\documentclass[10pt,conference]{IEEEtran}
\IEEEoverridecommandlockouts

\usepackage{cite}
\usepackage{amsmath,amssymb,amsfonts}
\usepackage{algorithmic}
\usepackage{graphicx}
\usepackage{textcomp}
\usepackage{color}
\def\BibTeX{{\rm B\kern-.05em{\sc i\kern-.025em b}\kern-.08em
    T\kern-.1667em\lower.7ex\hbox{E}\kern-.125emX}}

\newcommand{\cmd}[1]{\texttt{\color{blue!65!black}#1}}
\usepackage[scaled=0.95]{inconsolata}

\newcommand{\tool}{\textsc{Caprese}\xspace}

\newcommand{\js}{{Java\-Script}\xspace}
\newcommand{\numberOfSubjects}{{10}\xspace}

\usepackage{booktabs}

\usepackage{algorithmic}
\usepackage{algorithm}
\usepackage{listings}
\usepackage{color}
\usepackage{multirow}
\usepackage{array}
\usepackage{xspace}
\usepackage{hyperref}
\usepackage[utf8]{inputenc}
\usepackage{balance}
\usepackage{arydshln}
\usepackage{amsmath}
\usepackage{soul}  
\setulcolor{red}

\definecolor{gray}{rgb}{0.5,0.5,0.5}
\definecolor{mauve}{rgb}{0.58,0,0.82}
\definecolor{oliveGreen}{rgb}{0.5,.75,0.32}

\lstdefinelanguage{JavaScript}{
  morekeywords=[1]{break, continue, delete, else, for, function, if, in,
    new, return, this, typeof, var, void, while, with},
  morekeywords=[2]{false, null, true, boolean, number, undefined,
    Array, Boolean, Date, Math, Number, String, Object},
  morekeywords=[3]{eval, parseInt, parseFloat, escape, unescape},
  sensitive,
  morecomment=[s]{/*}{*/},
  morecomment=[l]//,
  morecomment=[s]{/**}{*/}, 
  morestring=[b]',
  morestring=[b]"
}[keywords, comments, strings]

\usepackage{cite}
\usepackage{amsmath,amssymb,amsfonts}
\usepackage{algorithmic}
\usepackage{graphicx}
\usepackage{textcomp}
\usepackage{xcolor}
\def\BibTeX{{\rm B\kern-.05em{\sc i\kern-.025em b}\kern-.08em
    T\kern-.1667em\lower.7ex\hbox{E}\kern-.125emX}}
\begin{document}

\title{Caprese: Change Impact Recommendation Using History and Runtime Analysis}

\author{\IEEEauthorblockN{Sadjad Tavakoli}
\IEEEauthorblockA{ 
\textit{Magellan AI}\\
BC, Canada \\
sadjad\_tavakoli@sfu.ca}
\and
\IEEEauthorblockN{Saba Alimadadi}
\IEEEauthorblockA{ 
\textit{Simon Fraser University}\\
BC, Canada \\
saba@sfu.ca}
}

\maketitle

\begin{abstract}
Change impact recommendation identifies program
entities potentially affected by source code changes.
Historical co-changes and runtime dependencies provide
complementary evidence, particularly in JavaScript,
where dynamic and asynchronous behaviour complicates
dependency analysis.

We present \tool, an open-source framework for
function-level change impact recommendation in Node.js.
Given a change set, it produces a ranked impact set
of potentially affected functions.
It supports repository-based recommendation through
confidence-guided mining of recurring, change-set-relevant
co-change patterns of varying lengths; dynamic
recommendation through execution-derived dependencies
involving calls, callbacks, events, timers, and shared
state; and hybrid recommendation integrating both
analyses. Its modular implementation supports
reusable analyses.

Evaluation on ten open-source applications and 100
selected change sets shows that, on average across
projects, 77.4\% of the relevant functions among
the top 60 hybrid recommendations are identified
exclusively by one analysis, demonstrating substantial
complementarity. Dynamic recommendation achieves
the highest mean precision among the top five
recommendations (0.65), while repository-based
recommendation performs best when the ranking is
extended to 60 recommendations (0.42).
Hybrid recommendation outperforms TARMAQ
on average but does not consistently outperform
the individual modes. \tool is publicly available
with source code, documentation, and evaluation
artifacts.
\end{abstract}



\section{Introduction}
\label{sec:introduction}

Understanding the downstream effects of source code
changes is essential to software maintenance.
Given a \emph{change set} of modified program entities,
change impact analysis identifies an \emph{impact set}
of potentially affected entities, while change impact
recommendation prioritizes them for inspection,
modification, or regression testing~
\cite{theFirstImpactanalysis1996,Chianti2004,
ImpactMiner2012,domSensitive2015}.

Repository-based recommendation identifies
relationships through historical co-changes,
but may include incidental changes or miss
dependencies absent from repository history~
\cite{Zimmermann2005,TARMAQ2016}.
In \js, dynamic dispatch, callbacks, and
asynchronous execution complicate static
dependency analysis~
\cite{sabaUnderstanding2014,domSensitive2015,
FrankTip2013}.
Dynamic analysis captures execution-derived
dependencies but is limited to observed
executions. Integrating historical and runtime
evidence can therefore provide complementary
coverage of potentially impacted functions.

We present \tool, an open-source framework
for function-level change impact recommendation
in Node.js applications. Given a change set
of modified functions, \tool produces a ranked
impact set, termed the \emph{Candidate Inspection
Set} (CIS), through three modes:
\emph{repository-based recommendation}, using
confidence-guided mining of recurring,
change-set-relevant co-change patterns of
varying lengths;
\emph{dynamic recommendation}, capturing
dependencies involving calls, callbacks, events,
timers, and shared state; and
\emph{hybrid recommendation}, combining both.
Its modular implementation supports reuse
of historical and runtime analyses across
change sets.

We evaluate \tool on \numberOfSubjects\
open-source Node.js applications using 100
selected change sets and manually established
reference impact sets. On average across
projects, 77.4\% of the relevant functions
among the top 60 hybrid recommendations
are identified exclusively by one analysis,
demonstrating substantial complementarity.
Dynamic recommendation achieves the highest
mean precision at Top-5 and Top-20, while
repository-based recommendation leads at
Top-60. Hybrid recommendation outperforms
TARMAQ~\cite{TARMAQ2016} on average,
but does not consistently outperform the
individual modes.

The main contributions are:
\begin{itemize}
    \item \tool, an open-source Node.js
    impact recommendation framework integrating
    targeted co-change pattern mining, runtime
    dependency analysis, and hybrid ranking,
    with reusable components and documentation.
    \footnote{\url{https://github.com/SEatSFU/caprese}}

    \item An empirical evaluation and
    reusable artifacts characterizing
    recommendation precision, signal
    complementarity, and limitations across
    ten open-source applications.
\end{itemize}

\section{Caprese: Design and Implementation}
\label{sec:approach}

\subsection{Architecture and Workflow}
\label{subsec:architecture}

\begin{figure}[t]
    \centering
    \includegraphics[width=\columnwidth]
        {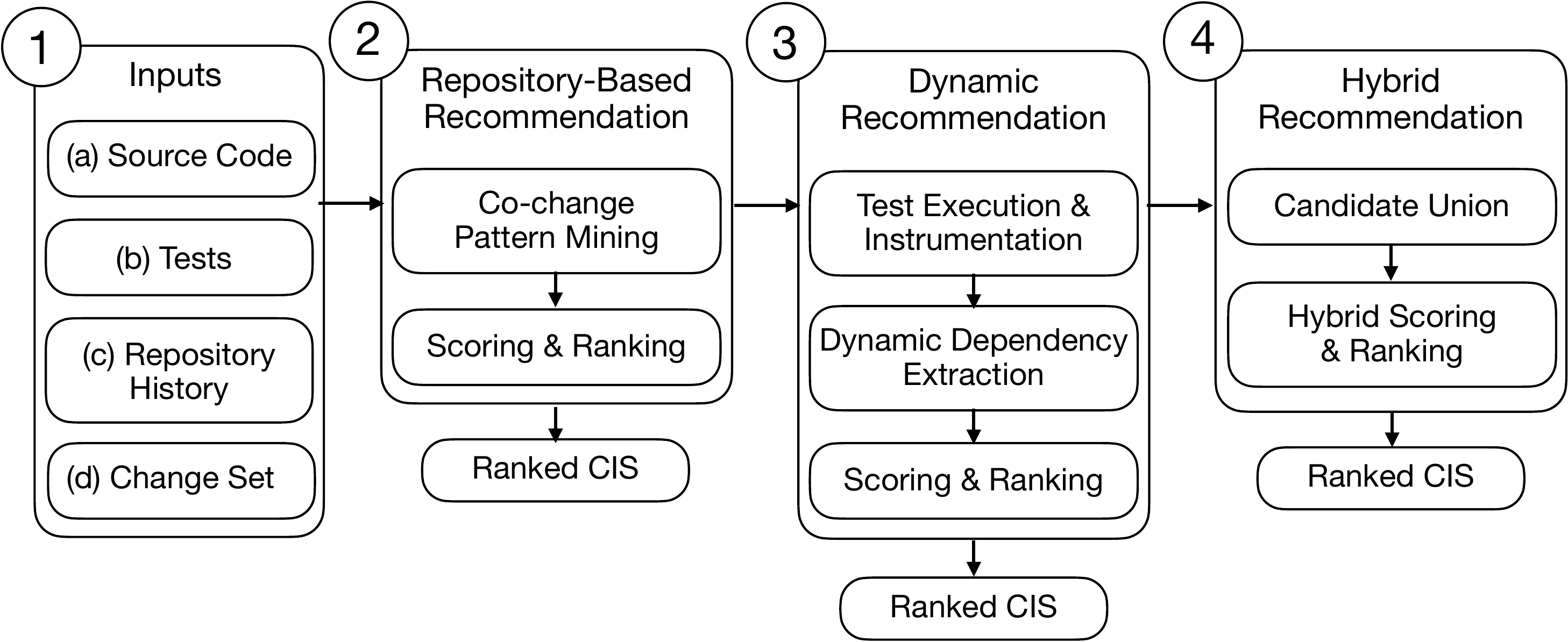}
    \caption{\tool's workflow: repository-based and
    dynamic analyses generate ranked CISs, which
    the hybrid mode combines.}
    \label{fig:overview}
\end{figure}

\autoref{fig:overview} illustrates \tool, an automated
framework for function-level change impact recommendation
in \js.
Given a change set of modified functions,
source code, tests, and repository history, \tool
produces a ranked Candidate Inspection Set (CIS)
of functions potentially impacted by the changes.
It supports three modes: repository-based recommendation
from historical co-change patterns, dynamic recommendation
from execution-derived dependencies, and hybrid
recommendation combining both sources of evidence.

\subsection{Repository-Based Recommendation}
\label{subsec:history}

The repository-based mode analyzes \emph{evolutionary
coupling}: functions that repeatedly change together
may require joint inspection.
Unlike transaction-overlap approaches such as
TARMAQ~\cite{TARMAQ2016}, \tool independently
mines recurring function-level co-change
patterns associated with the input change set.

\textbf{History extraction.}
\tool represents each commit as a transaction of modified
functions. It uses RefDiff's~\cite{Refdiff2020} code structure trees (CSTs)
to identify changes between consecutive revisions and
extends this analysis to map the detected changes to
their enclosing functions.

\textbf{Targeted pattern mining.}
\tool retains transactions overlapping the change set
and identifies co-occurring functions within them.
Starting with single-function patterns, it recursively
extends patterns using observed co-changes, prioritizing
extensions that may reveal new inspection candidates.
Unlike pairwise co-change analysis, this search discovers
patterns of varying lengths that overlap with subsets
of the change set, without enumerating all patterns
across the repository.

For a pattern $P$, $\mathrm{support}(P)$ is the number of
transactions containing all its functions. Confidence
measures how frequently the complete pattern occurs
relative to its change-set functions:
\begin{equation}
\label{eq:confidence}
\mathrm{confidence}(P)=
\frac{\mathrm{support}(P)}{\mathrm{support}(P_c)},
\end{equation}
where $P_c$ contains the functions of $P$ belonging to
the change set. This normalization directs mining
toward patterns associated with the current changes,
rather than merely frequent patterns in the repository.
Patterns meeting a configurable minimum confidence
threshold contribute candidates to the CIS. For each
candidate, \tool retains the confidence and support
of its highest-confidence pattern for ranking.

\subsection{Dynamic Dependency Recommendation}
\label{subsec:dynamic}

The dynamic mode instruments the application using
NodeProf\footnote{
\url{https://www.dag.inf.usi.ch/software/nodeprof}}
and executes its tests to collect runtime traces.
These traces are processed into a directed dependency
graph whose vertices represent functions, with special
handling for file-level execution and classes.

\textbf{Dependency extraction.}
\tool infers dependencies from function calls and
callbacks, event registration and emission, timers,
and reads and writes to shared state. Edges represent
possible directions of change propagation rather
than merely call directions. For example, if
function $f$ invokes $g$, \tool adds an edge from
$g$ to $f$, since a change to $g$ may affect its
caller. Additional edges capture argument passing,
event-related interactions, deferred callbacks,
and shared-state dependencies.

These relationships capture indirect interactions
beyond conventional call graphs. For example,
a function emitting an event may be related to
another function registering its listener,
even without a direct invocation between them.

\textbf{Candidate generation and ranking.}
Starting from the changed functions, \tool traverses
the dependency graph in breadth-first order and
collects reachable functions as inspection candidates.
Candidates are ranked by their shortest dependency
distance from the change set. Dependencies not
exercised by the available tests remain undetected.

\subsection{Hybrid Recommendation}
\label{subsec:hybrid}

The hybrid mode combines candidates from both analyses,
prioritizing functions identified by both. Shared
candidates are ranked by increasing runtime dependency
distance, breaking ties by decreasing historical
support and then confidence. Dynamic-only candidates
are ranked by dependency distance, while history-only
candidates are ranked by support and then confidence.
The two exclusive lists are interleaved after the
shared candidates, preserving recommendations unique
to either analysis.

\begin{table*}[t]
\centering
\caption{Benchmark characteristics and contribution of each analysis to the highest-ranked relevant recommendations (\%). Repo/Dyn: exclusive to each component; Both: identified by both. Cutoff $K$ selects the top-$K$ recommendations
in the hybrid ranking.}
\label{tab:signalComplementarity}
\scriptsize
\renewcommand{\arraystretch}{1.1}
\begin{tabular}{lrrr|rrr|rrr|rrr}
\toprule
\multirow{2}{*}{\textbf{Project}} &
\multirow{2}{*}{\textbf{KLOC}} &
\multirow{2}{*}{\textbf{Commits}} &
\multirow{2}{*}{\textbf{Txns}} &
\multicolumn{3}{c|}{\textbf{Top-5}} &
\multicolumn{3}{c|}{\textbf{Top-20}} &
\multicolumn{3}{c}{\textbf{Top-60}} \\
\cmidrule(lr){5-7}
\cmidrule(lr){8-10}
\cmidrule(lr){11-13}
& & & &
\textbf{Repo} &
\textbf{Dyn} &
\textbf{Both} &
\textbf{Repo} &
\textbf{Dyn} &
\textbf{Both} &
\textbf{Repo} &
\textbf{Dyn} &
\textbf{Both} \\
\midrule
bignumber.js & 48 & 220 & 111 & 5.0 & 30.0 & 65.0 & 3.8 & 35.2 & 61.0 & 20.1 & 62.1 & 17.8 \\
session & 4 & 678 & 257 & 12.0 & 11.5 & 76.5 & 16.1 & 29.7 & 54.2 & 20.9 & 44.5 & 34.6 \\
jhipster-uml & 11 & 868 & 448 & 14.8 & 24.1 & 61.1 & 32.3 & 39.0 & 28.7 & 31.4 & 49.3 & 19.3 \\
grant & 11 & 1,134 & 567 & 43.8 & 6.2 & 50.0 & 52.6 & 21.9 & 25.5 & 53.3 & 35.0 & 11.7 \\
environment & 33 & 1,165 & 661 & 7.8 & 24.8 & 67.4 & 13.6 & 27.7 & 58.7 & 17.1 & 51.2 & 31.7 \\
cla-assistant & 46 & 1,177 & 634 & 44.4 & 5.6 & 50.0 & 54.8 & 18.8 & 26.4 & 60.4 & 19.9 & 19.7 \\
assemble & 26 & 1,662 & 926 & 72.2 & 20.4 & 7.4 & 68.7 & 28.9 & 2.5 & 64.3 & 34.2 & 1.5 \\
nock & 36 & 1,756 & 882 & 0.0 & 22.5 & 77.5 & 8.0 & 38.2 & 53.8 & 8.8 & 53.2 & 38.0 \\
fastify & 49 & 2,682 & 1,282 & 14.3 & 0.0 & 85.7 & 29.0 & 20.8 & 50.2 & 22.2 & 49.8 & 28.0 \\
express & 19 & 5,248 & 3,111 & 35.7 & 7.1 & 57.1 & 45.2 & 20.3 & 34.4 & 46.6 & 29.6 & 23.8 \\
\midrule
\textbf{Average}
& -- & -- & -- & \textbf{25.0} & \textbf{15.2} & \textbf{59.8} & \textbf{32.4} & \textbf{28.0} & \textbf{39.5} & \textbf{34.5} & \textbf{42.9} & \textbf{22.6} \\
\bottomrule
\end{tabular}
\end{table*}

\section{Tool Usage and Reproducibility}
\label{sec:usage}

\textbf{Implementation and organization.}
\tool implements three components: frequent pattern
detection (FPD), dynamic analysis (DA), and hybrid
recommendation. FPD extends
RefDiff\footnote{\url{https://github.com/aserg-ufmg/RefDiff}}
by mapping detected
source-code changes to their enclosing functions
across Git revisions. It implements change-set-guided
evolutionary coupling analysis inspired by sequential
pattern mining and CLASP~\cite{clasp}, with support
from SPMF~\cite{SPMF}, an open-source data mining
library.
DA uses NodeProf to instrument Node.js applications
and processes execution traces to construct a
directed dependency graph.
The hybrid component ranks candidates using
historical and runtime evidence, while the
\texttt{caprese.js} CLI coordinates the analyses
and produces ranked recommendations.
Historical and runtime analyses can be reused
across change sets.

\textbf{Installation and configuration.}
\tool requires Node.js, NodeProf,
Graal.js,\footnote{\url{https://www.graalvm.org/javascript/docs/}}
 Java,
Git,\footnote{\url{https://git-scm.com}}
and Maven;\footnote{\url{https://maven.apache.org}}
the modified RefDiff component
is built with Gradle.\footnote{\url{https://gradle.org}}
The repository provides
installation instructions and a Dockerfile.
Users specify the target repository, branch,
test directories, excluded paths, starting commit,
and data storage location in \texttt{constants.js}.

\textbf{Execution and output.}
Users invoke \cmd{node caprese.js mine} to
prepare historical co-change information and
\cmd{node caprese.js da} to collect runtime
dependencies through instrumented application
execution. They then run
\cmd{node caprese.js detect "..."} to generate
ranked recommendations for a specified change set:

\begin{quote}
\small\ttfamily
node caprese.js detect\\
"getAssist-main/assist/getAssists.js-12-18"
\end{quote}

Changed functions are identified by name, file path,
and source-line range; multiple functions can be
specified as space-separated entries. Recommendations
are written to \texttt{capreseResult.json} for
inspection or further analysis.

\subsection{Data and Tool Availability}
\label{subsec:availability}

\tool is available under the MIT license at
\url{https://github.com/SEatSFU/caprese},
with a versioned snapshot archived on
Zenodo~\cite{CapreseZenodo}. The repository includes
source code, documentation, analysis components,
evaluation scripts, and experimental data to
support reproducibility and further experimentation.

\section{Evaluation and Findings}
\label{sec:evaluation}

\subsection{Evaluation Setup}
\label{subsec:eval-setup}

We evaluated \tool on ten open-source Node.js
applications spanning 4--49 KLOC, 220--5,248 commits,
and 111--3,111 function-level change transactions.
We randomly selected ten distinct function-level
change sets per project (100 total), using all
remaining commits as historical data, including
commits that postdate the selected changes.
Runtime dependencies
were collected by instrumenting application code
with NodeProf and executing its tests.
\autoref{tab:signalComplementarity} summarizes
the benchmark characteristics.

Two investigators, including one non-author,
manually established reference impact sets (RISs)
using predefined JavaScript-specific guidelines
covering compile-time, runtime, transitive, and
artifact-mediated dependencies. Their assessments
were jointly confirmed, although the RISs may
not be exhaustive.

We compare \tool's repository-based, dynamic,
and hybrid modes with a function-level
reimplementation of TARMAQ~\cite{TARMAQ2016},
a representative evolutionary coupling-based
change recommendation technique.
Precision@$K$ measures the fraction of the top-$K$
recommendations included in the RIS, for
$K\in\{5,20,60\}$. At the same cutoffs,
complementarity measures the proportions of
relevant recommendations identified exclusively
by repository analysis, exclusively by dynamic
analysis, or by both.

\subsection{Complementarity of Recommendation Signals}
\label{subsec:eval-complementarity}

At Top-60, averaging across projects, 22.6\%
of relevant recommendations are identified by
both analyses, while 34.5\% are exclusive to
repository analysis and 42.9\% to dynamic
analysis (\autoref{tab:signalComplementarity}).
Thus, 77.4\% of relevant recommendations
within this cutoff are identified by only
one analysis, demonstrating substantial
complementarity between historical and
runtime evidence.

The overlap decreases from 59.8\% at Top-5
to 39.5\% at Top-20 and 22.6\% at Top-60.
Although higher overlap at smaller cutoffs
partly reflects the hybrid ranking's prioritization
of shared candidates, the low overlap at Top-60
shows that both analyses contribute substantial
distinct relevant candidates beyond the
highest-ranked functions.

Contributions also vary across projects.
At Top-60, repository-only candidates account
for 64.3\% in \emph{assemble} and 53.3\%
in \emph{grant}, whereas dynamic-only candidates
account for 62.1\% in \emph{bignumber.js}
and 53.2\% in \emph{nock}. This variation
further supports retaining both sources
of evidence.

\subsection{Recommendation Precision}
\label{subsec:eval-precision}

\begin{table*}[t]
\centering
\caption{Precision@$K$ of the recommendation approaches at three inspection budgets.}
\label{tab:precisionComparison}
\scriptsize
\renewcommand{\arraystretch}{1.1}
\begin{tabular}{lrrrr|rrrr|rrrr}
\toprule
\multirow{2}{*}{\textbf{Project}} &
\multicolumn{4}{c|}{\textbf{Top-5}} &
\multicolumn{4}{c|}{\textbf{Top-20}} &
\multicolumn{4}{c}{\textbf{Top-60}} \\
\cmidrule(lr){2-5}
\cmidrule(lr){6-9}
\cmidrule(lr){10-13}
&
\textbf{TARMAQ} &
\textbf{Repo} &
\textbf{Dynamic} &
\textbf{Hybrid} &
\textbf{TARMAQ} &
\textbf{Repo} &
\textbf{Dynamic} &
\textbf{Hybrid} &
\textbf{TARMAQ} &
\textbf{Repo} &
\textbf{Dynamic} &
\textbf{Hybrid} \\
\midrule
bignumber.js & 0.39 & 0.52 & 0.92 & 0.78 & 0.42 & 0.46 & 0.88 & 0.68 & 0.34 & 0.49 & 0.57 & 0.60 \\
session & 0.56 & 0.69 & 0.89 & 0.76 & 0.38 & 0.60 & 0.59 & 0.61 & 0.30 & 0.58 & 0.30 & 0.37 \\
jhipster-uml & 0.22 & 0.33 & 0.55 & 0.34 & 0.14 & 0.25 & 0.26 & 0.26 & 0.07 & 0.25 & 0.12 & 0.14 \\
grant & 0.42 & 0.53 & 0.37 & 0.51 & 0.22 & 0.47 & 0.39 & 0.44 & 0.18 & 0.46 & 0.22 & 0.32 \\
environment & 0.20 & 0.36 & 0.67 & 0.62 & 0.26 & 0.36 & 0.57 & 0.43 & 0.28 & 0.31 & 0.37 & 0.36 \\
cla-assistant & 0.30 & 0.36 & 0.80 & 0.48 & 0.21 & 0.33 & 0.43 & 0.37 & 0.19 & 0.31 & 0.24 & 0.27 \\
assemble & 0.40 & 0.68 & 0.65 & 0.71 & 0.24 & 0.64 & 0.46 & 0.66 & 0.22 & 0.64 & 0.47 & 0.64 \\
nock & 0.36 & 0.53 & 0.82 & 0.70 & 0.33 & 0.52 & 0.59 & 0.61 & 0.30 & 0.50 & 0.31 & 0.31 \\
fastify & 0.10 & 0.33 & 0.33 & 0.38 & 0.09 & 0.19 & 0.23 & 0.20 & 0.09 & 0.16 & 0.19 & 0.16 \\
express & 0.38 & 0.73 & 0.49 & 0.44 & 0.23 & 0.52 & 0.29 & 0.34 & 0.18 & 0.45 & 0.11 & 0.25 \\
\midrule
\textbf{Average}
& \textbf{0.33} & \textbf{0.51} & \textbf{0.65} & \textbf{0.57} & \textbf{0.25} & \textbf{0.43} & \textbf{0.47} & \textbf{0.46} & \textbf{0.22} & \textbf{0.42} & \textbf{0.29} & \textbf{0.34} \\
\bottomrule
\end{tabular}
\end{table*}

As shown in \autoref{tab:precisionComparison},
dynamic recommendation achieves the highest
mean Precision@5 (0.65) and Precision@20 (0.47),
while repository-based recommendation leads
at Precision@60 (0.42).
The strong performance of dynamic recommendation
at smaller cutoffs is consistent with the
usefulness of execution-derived dependencies
and proximity-based ranking. However, the
evaluation does not isolate the contributions
of individual dependency types.

Repository-based recommendation also exceeds
TARMAQ's mean precision at all three cutoffs.
Unlike TARMAQ's maximum transaction-overlap
strategy, \tool explores recurring multi-function
co-change patterns associated with the change set,
using confidence-guided mining and ranking.
These results demonstrate improved recommendation
precision over the reimplemented baseline,
without isolating the contributions of individual
algorithmic choices.

Hybrid recommendation exceeds TARMAQ's mean
precision at every cutoff (0.57 vs.\ 0.33 at
Top-5, 0.46 vs.\ 0.25 at Top-20, and 0.34 vs.\
0.22 at Top-60), but does not consistently
outperform the individual modes. Thus, while
integrating historical and runtime evidence
retains complementary relevant candidates,
the current hybrid ranking does not always
prioritize them as effectively as the
stronger individual mode.

\subsection{Practical Insights and Limitations}
\label{subsec:eval-insights}

Manual examination identified relevant coupling
captured through history but missed by runtime
analysis, including duplicated or semantically
similar code, dependencies mediated by external
resources, shared identifiers, and unexecuted
code. Conversely, historical analysis may
introduce false positives through incidental
co-changes, obsolete relationships, or
bidirectional coupling that obscures asymmetric
dependencies. Co-change does not necessarily
imply impact propagation. These observations
help explain the complementary strengths and
limitations of the two analyses.

\textbf{Execution cost.}
On an Apple M1 system, \tool generated ranked
recommendations in 3.09 seconds on average,
excluding dynamic dependency extraction,
compared with 2.66 seconds for TARMAQ.
Dynamic dependency extraction, including test
execution, required 95.47 seconds on average
and can be reused across change sets.

\textbf{Threats to validity.}
The ten applications and selected change sets
limit generalizability. Manually constructed RISs
may be subjective or incomplete despite predefined
guidelines and assessment by two investigators.
Using commits that postdate the selected changes
may overestimate performance under prospective
deployment.
The reported differences are descriptive;
statistical significance was not assessed.

\section{Related Work}
\label{sec:related-work}

\textbf{Dependency-based impact analysis.}
Prior techniques use static dependencies, program
slicing, and runtime relationships~
\cite{theFirstImpactanalysis1996,Chianti2004,
PathImpact02003,ImpSlicing2011}.
Static analysis faces challenges with JavaScript's
dynamic features, while dynamic analysis is
limited by execution coverage.
Tochal~\cite{domSensitive2015} combines static
and dynamic analysis for event-driven,
DOM-sensitive impact analysis in client-side
JavaScript. \tool instead targets Node.js
and combines runtime dependencies with
historical co-changes.
\textbf{Repository-based and hybrid recommendation.}
ROSE~\cite{Zimmermann2005} mines evolutionary
coupling, while TARMAQ~\cite{TARMAQ2016}
selects transactions with maximal change-set
overlap. Other approaches investigate transitive
associations~\cite{Islam2018}, adaptive history
selection~\cite{ATARI2020}, and sequential
co-change patterns~\cite{MiningSequencesKagdi2006}.
Hybrid approaches integrate history with textual,
structural, or execution evidence~
\cite{ImpactMiner2012,IntegratingConAndCoup2013,
interactionAndHistory2014}.
CoRec~\cite{Jiang2020} learns JavaScript
co-change recommendations from recurring
patterns and program features.
Unlike TARMAQ, \tool mines recurring
change-set-relevant function-level patterns
of varying lengths and integrates them
with runtime dependencies.

\section{Conclusion}
\label{sec:conclusion}

We presented \tool, an open-source framework
integrating function-level co-change pattern
mining and runtime dependency analysis for
Node.js change impact recommendation.
Evaluation on ten applications demonstrates
substantial complementarity between historical
and runtime evidence, although hybrid ranking
does not consistently outperform the individual
modes. The released tool and evaluation artifacts
support further research on impact recommendation,
including improved hybrid ranking, broader runtime
dependency coverage, and additional JavaScript
ecosystems.

\section*{Acknowledgments}

ChatGPT (OpenAI) was used to assist with reviewing and
revising a manuscript originally drafted by the
authors. The authors verified the revised content
and remain responsible for its accuracy and
conclusions. Generative AI was not used to
develop \tool, conduct experiments, or collect
experimental data.

\balance

\bibliographystyle{IEEEtran}
\bibliography{refs}

@String{Computing = "Computing" }

@String{Computer = "{IEEE} Computer" }

@String{Springer = "Springer-Verlag" }

@inproceedings{Refdiff2020,
  title={RefDiff: Detecting refactorings in version histories},
  author={Silva, Danilo and Valente, Marco Tulio},
  booktitle={2017 IEEE/ACM 14th International Conference on Mining Software Repositories (MSR)},
  pages={269--279},
  year={2017},
  organization={IEEE}
}

@inproceedings{MiningSequencesKagdi2006,
  title={Mining sequences of changed-files from version histories},
  author={Kagdi, Huzefa and Yusuf, Shehnaaz and Maletic, Jonathan I},
  booktitle={Proceedings of the 2006 international workshop on Mining software repositories},
  pages={47--53},
  year={2006}
}

@inproceedings{ImpactMiner2012,
  title={Integrated impact analysis for managing software changes},
  author={Gethers, Malcom and Dit, Bogdan and Kagdi, Huzefa and Poshyvanyk, Denys},
  booktitle={2012 34th International Conference on Software Engineering (ICSE)},
  pages={430--440},
  year={2012},
  organization={IEEE}
}

@inproceedings{Chianti2004,
  title={Chianti: a tool for change impact analysis of java programs},
  author={Ren, Xiaoxia and Shah, Fenil and Tip, Frank and Ryder, Barbara G and Chesley, Ophelia},
  booktitle={Proceedings of the 19th annual ACM SIGPLAN conference on Object-oriented programming, systems, languages, and applications},
  pages={432--448},
  year={2004},
  address = {New York, NY, USA},
  series = {OOPSLA '04}
}

@inproceedings{sabaUnderstanding2014,
  title={Understanding JavaScript event-based interactions},
  author={Alimadadi, Saba and Sequeira, Sheldon and Mesbah, Ali and Pattabiraman, Karthik},
  booktitle={Proceedings of the 36th International Conference on Software Engineering},
  pages={367--377},
  year={2014}
}

@article{Zimmermann2005,
  title={Mining version histories to guide software changes},
  author={Zimmermann, Thomas and Zeller, Andreas and Weissgerber, Peter and Diehl, Stephan},
  journal={IEEE Transactions on software engineering},
  volume={31},
  number={6},
  pages={429--445},
  year={2005},
  publisher={IEEE}
}

@inproceedings{TARMAQ2016,
  title={Generalizing the analysis of evolutionary coupling for software change impact analysis},
  author={Rolfsnes, Thomas and Di Alesio, Stefano and Behjati, Razieh and Moonen, Leon and Binkley, Dave W},
  booktitle={2016 IEEE 23rd International Conference on Software Analysis, Evolution, and Reengineering (SANER)},
  volume={1},
  pages={201--212},
  year={2016},
  organization={IEEE}
}

@inproceedings{Islam2018,
  title={Detecting evolutionary coupling using transitive association rules},
  author={Islam, Md Anaytul and Islam, Md Moksedul and Mondal, Manishankar and Roy, Banani and Roy, Chanchal K and Schneider, Kevin A},
  booktitle={2018 IEEE 18th International Working Conference on Source Code Analysis and Manipulation (SCAM)},
  pages={113--122},
  year={2018},
  organization={IEEE}
}

@inproceedings{domSensitive2015,
  title={Hybrid DOM-sensitive change impact analysis for JavaScript},
  author={Alimadadi, Saba and Mesbah, Ali and Pattabiraman, Karthik},
  booktitle={29th European Conference on Object-Oriented Programming (ECOOP 2015)},
  pages={321--345},
  year={2015}
}

@inproceedings{interactionAndHistory2014,
  title={Impact analysis of change requests on source code based on interaction and commit histories},
  author={Zanjani, Motahareh Bahrami and Swartzendruber, George and Kagdi, Huzefa},
  booktitle={Proceedings of the 11th Working Conference on Mining Software Repositories},
  pages={162--171},
  year={2014}
}

@inproceedings{clasp,
  title={Clasp: An efficient algorithm for mining frequent closed sequences},
  author={Gomariz, Antonio and Campos, Manuel and Marin, Roque and Goethals, Bart},
  booktitle={Pacific-Asia Conference on Knowledge Discovery and Data Mining},
  pages={50--61},
  year={2013},
  organization={Springer}
}

@article{nodeProf,
author = {Sun, Haiyang and Humer, Christian and Bonetta, Daniele and Binder, Walter},
journal = {CC 2018 - Proceedings of the 27th International Conference on Compiler Construction, Co-located with CGO 2018},
doi = {10.1145/3178372.3179527},
isbn = {9781450356442},
month = {feb},
pages = {196--206},
publisher = {Association for Computing Machinery, Inc},
title = {Efficient dynamic analysis for node.Js - NodeProf},
volume = {2018-Febru},
year = {2018}
}

@inproceedings{PathImpact02003,
  title={Incremental dynamic impact analysis for evolving software systems},
  author={Law, James and Rothermel, Gregg},
  booktitle={14th International Symposium on Software Reliability Engineering, 2003. ISSRE 2003.},
  pages={430--441},
  year={2003},
  organization={IEEE}
}

@book{theFirstImpactanalysis1996,
  title={Software change impact analysis},
  author={Arnold, Robert S},
  year={1996},
  publisher={IEEE Computer Society Press},
address = {Washington, DC, USA},
}

@inproceedings{ImpSlicing2011,
  title={Practical change impact analysis based on static program slicing for industrial software systems},
  author={Acharya, Mithun and Robinson, Brian},
  booktitle={Proceedings of the 33rd international conference on software engineering},
  pages={746--755},
  year={2011}
}

@article{ATARI2020,
  title={On adaptive change recommendation},
  author={Moonen, Leon and Binkley, David and Pugh, Sydney},
  journal={Journal of Systems and Software},
  volume={164},
  pages={110550},
  year={2020},
  publisher={Elsevier}
}

@article{IntegratingConAndCoup2013,
  title={Integrating conceptual and logical couplings for change impact analysis in software},
  author={Kagdi, Huzefa and Gethers, Malcom and Poshyvanyk, Denys},
  journal={Empirical Software Engineering},
  volume={18},
  number={5},
  pages={933--969},
  year={2013},
  publisher={Springer}
}

@article{Jiang2020,
  title={Investigating and recommending co-changed entities for JavaScript programs},
  author={Jiang, Zijian and Zhong, Hao and Meng, Na},
  journal={Journal of Systems and Software},
  volume={180},
  pages={111027},
  year={2021},
  publisher={Elsevier}
}

@article{SPMF,
  title={SPMF: a Java open-source pattern mining library},
  author={Fournier-Viger, Philippe and Gomariz, Antonio and Gueniche, Ted and Soltani, Azadeh and Wu, Cheng-Wei and Tseng, Vincent S},
  journal={The Journal of Machine Learning Research},
  volume={15},
  number={1},
  pages={3389--3393},
  year={2014},
  publisher={JMLR. org}
}

@inproceedings{FrankTip2013,
  title={Efficient construction of approximate call graphs for JavaScript IDE services},
  author={Feldthaus, Asger and Sch{\"a}fer, Max and Sridharan, Manu and Dolby, Julian and Tip, Frank},
  booktitle={2013 35th International conference on software engineering (ICSE)},
  pages={752--761},
  year={2013},
  organization={IEEE}
}

@software{CapreseZenodo,
  author    = {Tavakoli, Sadjad and Alimadadi, Saba},
  title     = {Caprese: Change Impact Recommendation
               Using History and Runtime Analysis},
  version   = {1.0.0},
  year      = {2026},
  publisher = {Zenodo},
  doi       = {10.5281/zenodo.24120003},
  url       = {https://doi.org/10.5281/zenodo.24120002}
}

\end{document}